
\magnification=\magstep1
\overfullrule=0pt
\def\P{\cal P}

\def\Buildrel#1\over#2{\mathrel{\mathop{\kern0pt #1}\limits_{#2}}}
\quad
\hfill
HD-THEP-95-35
\vskip1.5cm
\centerline{\bf LATTICE ENERGY SUM RULES AND THE TRACE ANOMALY}
\vskip1cm
\centerline{Heinz J. Rothe}
\medskip
\centerline{Institut f\"ur Theoretische Physik}
\centerline{Universit\"at Heidelberg}
\centerline{Philosophenweg 16, D-69120 Heidelberg}
\vskip2.5cm
\centerline{\bf Abstract}
\bigskip
\baselineskip21pt
\noindent
We show that the additional contribution to the Michael lattice energy sum
rule for the static quark-antiquark potential, pointed out recently, can
be
identified with the contribution to the
field energy arising from the trace anomaly
of the energy momentum tensor. We also exlicitely exhibit the anomalous
contribution to the field energy in the sum rule for the glueball mass
obtained recently by Michael.
\vfill\eject
\baselineskip21pt
Recently, there has been a renewed interest in the action and energy
lattice Michael sum rules [1], which relate the static $q\bar q$ potential
in an SU(N) gauge theory to the action and energy stored in the colour
electric and magnetic fields, after it had been pointed out in [2],
that an important contribution to the action sum rule had not been taken
into account in [1]. Subsequently it was found [3] that also the energy
sum rule obtained in [1] is modified in an
essential way by an additional contribution to the colour electric and
magnetic field energy which, in the case of a confining $q\bar
q$-potential,
accounts for
half of the energy in the flux tube matching the interquark potential.
That the derivation of the sum rule in [1] was incomplete has also
been noticed in [4], where the authors however speculate that the
additional contribution they find is an artefact of a non bonafide
approximation.
In [3] we had not given a physical interpretation of the additional
contribution appearing in the modified energy sum rule. This note is
intended to fill the gap. In particular we will show that the additional
term is nothing but the contribution to the field energy coming from the
trace anomaly [5] of
the energy momentum tensor. That the trace anomaly plays an important
role in constructing phenomenological hadron models, and accounts,
for example, for 1/4  of the nucleon mass, has been recently stressed in
[6]. This paper has been the trigger for the present note.
In lattice perturbation theory the trace anomaly has been computed in
one loop order by Caracciolo, Menotti and Pelisetto [7].
Starting from a traceless,
non-conserved tree-level energy momentum tensor on the lattice,
and correcting it by imposing the Ward identities for its conservation,
the authors show that the corrected energy momentum tensor reproduces
the correct anomaly in the continuum limit. This anomaly must
contribute to the field energy [6] matching the
$q\bar q$-potential, and should manifest itself in the lattice energy
sum rule. This we now show.

Starting from the expression for the expectation value of the Wilson loop
on an anisotropic lattice with temporal lattice spacing $a_\tau$ and
spatial lattice spacing $a$, and requiring that for
$a \to 0$, $a_\tau \to 0$, with $\xi = a/a_\tau$ fixed, the potential
be independent of the anisotropy parameter $\xi$,
we arrived in [3] at the following relation connecting the interquark
potential to correlators of plaquette variables with the Wilson loop,
$$\hat V(\hat R,\hat\beta)=\eta_- <-{\P}_\tau' + {\P}_s'>_{q\bar q-0}
-\eta_+<{\P}_\tau' + {\P}_s'>_{q\bar q-0} \ ,\eqno(1)$$
where ${\cal P}'_\tau$ and ${\cal P}'_s$ stand for the time-like and
space-like plaquette contributions to the action with base on a fixed time
slice (taken to be the $x_4 = 0$ plane), and $<{\cal O}>_{q\bar q -0}$
denotes generically the following correlator with the Wilson loop
$W(\hat R,\hat T)$
$$<O>_{q\bar q-0}={<W(\hat R,\hat T)O>\over<W(\hat R,\hat T)>}
-<O>.\eqno(2)$$
In the temporal direction the loop is taken to extend from $-{\hat T\over
2}$
to ${\hat T\over 2}$, with $\hat T$ very large. $\hat R$ is the separation
of the quark-antiquark pair. Quantities measured in lattice units are
denoted with a "hat". In (1) a subtraction at some reference
quark-antiquark
separation is implied, in order to elliminate the self energy
contributions.
The coefficients $\eta_{\pm}$ are related to the $\xi$-
derivatives of the couplings $\beta_\tau (a,\xi)$ and
$\beta_s (a,\xi)$ associated with the time-like
and space-like plaquette contributions, ${\cal P}_\tau$ and ${\cal P}_s$,
to the action on an anisotropic lattice,
$S = \beta_\tau{\cal P}_\tau+\beta_s{\cal P}_s$, by
 $\eta_{\pm}=(1/2)\left [(\partial\beta_\tau
/\partial\xi)
 \pm (\partial\beta_s/\partial\xi)\right]_{\xi=1}$,
where $\beta_\tau(a,1) = \beta_s(a,1) = \hat\beta (a)$, with $\hat\beta  =
2N/g^2_0$ the usual SU(N) gauge coupling on an isotropic lattice (for
details see ref. [3]).
We have denoted it with a "hat", in order to avoid any confusion with the
$\beta$-function. The correlators
in (1) are calculated with the Wilson action on an isotropic
lattice.
\smallskip
Consider the second term appearing on the rhs of (1), which is the
essential modification of the sum rule given in [1], referred to above.
The coefficient $\eta_+$ has been calculated non-perturbatively by Karsch
[8],
and can be written in the form
$$\eta_+ = -{\beta_L(g_0)\over 2g_0}\hat\beta \ , \eqno(3)$$
where $\beta_L(g_0) = -a{\partial g_0\over\partial a}$ is the lattice
$\beta$-function. Hence the contribution of the second term appearing
on the rhs of (1) is given by
$$-\eta_+ <{\cal P}'_\tau + {\cal P}'_s>_{q\bar q-0} =
{1\over 4}\left({{2\beta_L\over g_0}}<\hat L>_{q\bar q-0}\right)
 \ , \eqno(4a)$$
where
$$\hat L = \hat\beta ({\cal P}'_\tau + {\cal P}'_s)\eqno(4b)$$
is the (dimensionless) lattice version of the (euclidean) continuum
Lagrangian density integrated over all space at a fixed time.
The quantity $(2\beta_L/g_0)\hat L$ in (4a) has presicely the form of
the trace anomaly computed in lattice perturbation theory in ref. [7],
summed over the spatial lattice sites on a fixed time slice. Since the
(euclidean) energy momentum tensor can be decomposed into a traceless-and
trace part, i.e.,
$T_{\mu\nu} = (T_{\mu\nu} - {1\over 4}\delta_{\mu\nu}T) +
{1\over 4}\delta_{\mu\nu}T \ , $
where T denotes the (euclidean) trace of $T_{\mu\nu}$, we conclude that
the second term appearing on the rhs of (1) is the
contribution to the field energy of a $q\bar q$-pair (determined by the
space integral of $T_{44}$) arising from the trace anomaly. In the
weak coupling limit, corresponding to vanishing lattice spacing, the
anomalous contribution to the potential (1), in physical units,
takes the form
$$V_a(R,g_0(a),a) = {{\beta_L(g_0)\over g_0}}<\int d^3x {1\over 4}
[E^2(x) + B^2(x)]>_{q\bar q-0} \ , \eqno(5)$$
where $E^2$ and $B^2$ denote the square of the (euclidean) colour electric
and magnetic fields. A summation over colours is understood.
The subscript "a" on $V_a$ stands for "anomalous".
This is the lattice cutoff version of the contribution
of the trace anomaly to the gluon field energy discussed in [6],
generalized to
the presence of a quark-antiquark pair. For $a \to 0$, the expression (5)
is a finite, renormalization group invariant expression. By
following a similar line of arguments as in [2], it can be
expressed in terms of a renormalized coupling constant g, and renormalized
squared colour electric and magnetic fields. The form of the rhs of (5)
remains the same, except that $\beta_L(g_0)/g_0$ is replaced by $\beta
(g)/g$,
where $\beta (g) = \mu \partial g/\partial\mu$ is the continuum
beta-function, with $\mu$ the renormalization scale. The corresponding
anomalous contribution to the potential is thus given by 1/4 of the space
integral of the trace anomaly obtained in [5].

The contribution of the traceless part of the energy momentum tensor to
the field energy is expected to be given by the first term in (1).
This is consistent with the observation that in the weak coupling limit
($g_0 \to 0$), $\eta_- \to \hat\beta$ [8], and that for $a \to 0$, with
$R = \hat Ra$ fixed and the dependence of $\hat\beta (a)$ given by
the usual renormalization group relation,
$${1\over a}\hat\beta <-{\cal P}'_\tau + {\cal P}'_s>_{q\bar q-0}
\to <\int d^3x {1\over 2}(-E^2 + B^2)>_{q\bar q -0} \ .\eqno(6)$$
The quantity appearing within brackets is the usual expression for
the colour electric and magnetic field energy in the absence of
a trace anomaly, expressed in terms of the euclidean fields.
\smallskip
What concerns the lattice action sum rule obtained in [3,4]
(derived in [2] in the continuum formulation), it can be written in the
form
$$\hat V(\hat R,\hat\beta) +
\hat R{\partial \hat V(\hat R,\hat\beta)\over\hat R} =
{2\beta_L\over g_0}<\hat L>_{q\bar q-0}\ . \eqno(7)$$
The rhs of this equation is just the trace of the energy momentum
tensor summed over
the spatial lattice sites at fixed (euclidean) time. This is similar
to the observation made by Michael [1], that the action sum rule
for the lightest glueball mass relates this mass to the trace anomaly.
For a confining potential $\hat V = \hat\sigma\hat R$,
with $\hat\sigma$ the string tension in lattice units, the lhs
of (7) is just twice the potential. Hence the contribution to the field
energy in (1) arising from the trace anomaly is one-half of the interquark
potential.

The trace anomaly is, of course, also expected to manifest itself in the
energy sum rule for the mass of a glueball.
In a recent paper [9], Michael derived a lattice energy sum rule
for the mass of the lightest glueball with a
given set of quantum numbers, by departing from a lattice formulation of
QCD on an anisotropic lattice with unequal lattice spacings in all four
space-time directions. In our notation this sum rule takes the form
$$M = {4\over 3}\eta_-<-{\cal{P}}'_\tau + {\cal{P}}'_s>_{1-0}\ \ .
\eqno(8)$$
The bracket $<{\cal O}>_{1-0}$ stands generically for the following
correlator
$$ <{\cal O}>_{1-0} = {<{\cal O}G(\tau)>
\over <G(\tau)>} - <{\cal O}> \ , \eqno(9)$$
where $G(\tau)$ is an operator which excites glueball states in a
given representation of the symmetry group, and whose expectation value,
$<G(\tau)>$, decays like $\exp (-M\tau)$ for large euclidean times, with M
the lightest glueball mass. Such operator can be constructed from
products of Wilson loops located on two different time slices, separated
by a (euclidean) time interval $\tau$ [1]. In the weak coupling limit,
where
$\eta_- \to \hat\beta$, the rhs of (8) is four thirds of the
expectation value of the classical field energy $\int d^3x
{1\over 2}[E^2 + B^2]$ in the one-gluon state, measured
relative to the vacuum. From (8) Michael [9] concluded that
the naive, semiclassical expression for the field
energy surrounding a glueball
accounts for only ${3\over 4}$ of the glueball mass. The author attributed
this departure from the
naively expected semiclassical result to vacuum polarization
effects which give rise to an effective dielectric constant. A closer look
at the sum rules of ref. [9] shows that the remaining field energy
required to match the
glueball mass comes from the anomalous contribution to the Hamiltonian
arising from the trace anomaly. This can be
readily seen by making use of a further sum rule obtained in [9],
relating the correlators involving temporal-like and space-like
plaquettes, to express the rhs of (8) as a sum of a normal and anomalous
contribution to the field energy. This sum rule, which is just a linear
combination of the energy and action sum rules derived in [9], can be
written
in the form
$(1/3)\eta_-<{\cal{P}}'_\tau - {\cal{P}}'_s>_{1-0} =
\eta_+<{\cal{P}}'_\tau + {\cal{P}}'_s>_{1-0}$.
Making use of this relation, as well as of (3) (which is equivalent to
the consistency equation in ref. [9]), the rhs of (8) can be separated
into two contributions as follows:
$$\hat M = \eta_-<-{\cal{P}}'_\tau + {\cal{P}}'_s>_{1-0} +
{\beta_L(g_0)\over 2g_0}<\hat L>_{1-0}\ , \eqno(10)$$
where $\hat L$ has been defined in (4b).
The first term on the rhs is just 3/4 of the glueball mass, as follows
from (8), and corresponds in the weak coupling limit to the
semiclassical field energy surrounding the glueball. The second term is
the contribution to the field energy arising from the trace
anomaly, which provides 1/4 of the glueball mass. The rhs of (10) has
exactly
the same form as that for the
static $q\bar q$- potential (1) with $\eta_+$ replaced by (3).
In fact it can be also readily
derived by proceeding along the lines of ref. [3], with the Wilson loop
replaced by the operator $G(\tau)$. While in the case of a
confining $q\bar q$-potential the trace anomaly surplied 1/2 of the
field energy stored in the flux tube, it only accounts for 1/4 of the
glueball
mass. The origin for this different
behaviour can be traced to the dependence of
the potential on the quark-antiquark separation which
leads to different action sum rules in the two cases. Thus for the
glueball,
the action sum rule obtained in [9] can be put into the form
$$\hat M = {2\beta_L\over g_0}<\hat L>_{1-0}\ , $$
while for a linear $q\bar q$-potential, the lhs of the action sum rule
(7) is twice the potential.

Concluding, we have seen that by extracting the $q\bar q$-potential
from the expectation value of the Wilson loop on an
anisotropic lattice, as described in [3], one is led in a straight
forward way to an energy sum rule, in which the contribution to the field
energy arising from the
traceless and trace parts of the energy momentum tensor can be readily
identified. Furthermore, we have seen that a similar sum rule holds
for the glueball mass, except that in this case the plaquette variables
are
correlated with the operator $G(\tau)$ whose exponential decay for large
euclidean times determines the lightest glueball mass.
\bigskip\bigskip
We are very grateful to T. Reisz and to O. Nachtmann for having  called
our attention to ref. [7] and to the recent preprint of X. Ji, ref. [6].
\vfill\eject
\centerline{\bf REFERENCES}
\bigskip\noindent
[1] C. Michael, Nucl. Phys. {\bf B280} [FS18] (1987) 13
\smallskip\noindent
[2] H.G. Dosch, O. Nachtmann and M. Rueter, Heidelberg preprint HD-THEP-
95-12, hep-ph 9503386
\smallskip\noindent
[3] H.J. Rothe, Phys. Lett. {\bf B355} (1995) 101
\smallskip\noindent
[4] G.S. Bali, K. Schilling and Ch. Schlichter, Phys. Rev. {\bf D51}
(1995) 5165
\smallskip\noindent
[5]  J.C. Collins, A. Duncan and S.D. Joglekar, Phys. Rev. {\bf D16}
(1977) 438;
\smallskip\noindent
[6] X. Ji, "Breakup of Hadron Masses and Energy-Momentum Tensor of QCD",
MIT preprint, MIT-CTP-2407 (1995)
\smallskip\noindent
[7] S. Caracciolo, P. Menotti and A. Pelissetto, Nucl. Phys. {\bf B375}
(1992) 195
\smallskip\noindent
[8] F. Karsch, Nucl. Phys. {\bf B205} [FS5] (1982) 285
\smallskip\noindent
[9] C. Michael, Lattice Sum Rules for Coloured Fields,
Liverpool preprint LTH 348,
hep-lat/9504016 (1995)
\end